\documentclass[12pt]{article}
\usepackage{amsmath,amssymb}

\usepackage{graphicx}
\usepackage{wrapfig}

\usepackage{hyperref}
\usepackage{cite}

\def\beq{\begin{eqnarray}}
\def\eeq{\end{eqnarray}}

\def\la{\langle }
\def\ra{\rangle }

\newcommand{\Tr}{\,\mathrm{Tr}\,}            

\newcommand{\be}{\begin{equation}}
\newcommand{\ee}{\end{equation}}
\newcommand{\bea}{\begin{eqnarray}}
\newcommand{\eea}{\end{eqnarray}}
\newcommand{\bg}{\begin{gather}}

\newcommand{\bseq}{\begin{subequations}}
\newcommand{\eseq}{\end{subequations}}

\renewcommand{\ln}{\mathop{\rm ln}\nolimits}

\def\half{\frac{1}{2}}

\def\tr{\hbox{Tr}}

\def\be{\begin{eqnarray}}
\def\ee{\end{eqnarray}}
\def\lb{\label}

\begin{document}

\title{\textbf{Entropy vs Gravitational Action: \\
Do Total Derivatives Matter?   }}

\vspace{2cm}
\author{ \textbf{  Amin Faraji Astaneh$^{1}$, Alexander Patrushev$^{2,3}$ and  Sergey N. Solodukhin$^3$ }} 
\date{}
\maketitle
\begin{center}
\hspace{-0mm}
  \emph{$^1$ Department of Physics, Sharif University of Technology,}\\
\emph{P.O. Box 11365-9161, Tehran, Iran\\
 and  School of Particles and Accelerators,\\ Institute for Research in Fundamental Sciences (IPM),}\\
\emph{ P.O. Box 19395-5531, Tehran, Iran}
 \end{center}
 \begin{center}
  \hspace{-0mm}
  \emph{ $^2$ Bogoliubov Laboratory of Theoretical Physics,}\\
  \emph{Joint Institute for Nuclear Research,}\\
  \emph{6 Joliot-Curie, 141980 Dubna, Russia}
\end{center}
\begin{center}
  \hspace{-0mm}
  \emph{ $^3$ Laboratoire de Math\'ematiques et Physique Th\'eorique  CNRS-UMR
7350, }\\
  \emph{F\'ed\'eration Denis Poisson, Universit\'e Fran\c cois-Rabelais Tours,  }\\
  \emph{Parc de Grandmont, 37200 Tours, France}
\end{center}



\begin{abstract}
\noindent { The total derivatives in the gravitational action are usually disregarded as  non-producing any non-trivial dynamics.
In the context of the gravitational entropy, within Wald's approach,  these terms are considered irrelevant as non-contributing to the entropy. 
On the other hand, the total derivatives are usually present in the trace anomaly in dimensions higher than 2. 
As the trace anomaly is related to the logarithmic term in the entanglement entropy it is natural to ask whether the total derivatives
make any essential contribution to the entropy or they can be totally ignored. In this  note we analyze this question for some particular examples
of total derivatives. Rather surprisingly, in all cases  that we consider the total derivatives   produce  non-trivial contributions to
the entropy.  Some of them are non-vanishing even if the extrinsic curvature of the surface is zero. 
We suggest that this may  explain the earlier observed discrepancy between the holographic entanglement entropy and Wald's entropy.
}
\end{abstract}

\vskip 1 cm
\noindent
\rule{7.7 cm}{.5 pt}\\
\noindent 
\noindent
\noindent ~~~ {\footnotesize e-mails: faraji@ipm.ir,   apatrush@gmail.com,  Sergey.Solodukhin@lmpt.univ-tours.fr}

\newpage
    \tableofcontents
\pagebreak

\newpage

\section{ Introduction}
Since the inspiring paper of Wald \cite{Wald:1993nt} and the subsequent works  \cite{Jacobson:1993xs}, \cite{Jacobson:1993vj} it became clear that there exists a certain correspondence 
between  the entropy associated to horizons and 
terms in the gravitational action. This relation by now is very well established and has many important applications and generalizations, see for instance \cite{Fursaev:1995ef} and 
\cite{Lewkowycz:2013nqa}. 
In this context the possible total derivatives in the gravitational action are generally neglected as they are thought not to  produce any essential contribution to the entropy.

In a wider context the discussed  correspondence is important for the calculation of entanglement entropy of a co-dimension two surface. 
The gravitational action in this case is the quantum effective action which in general can be represented as a certain local or non-local expansion \cite{Barvinsky:1985an} in Riemann 
curvature.  In particular, there has been established  \cite{Ryu:2006ef}, \cite{Solodukhin:2008dh} a relation between the trace anomaly (obtained as variation of the effective action under conformal rescaling of metric) and the logarithmic term in the entanglement entropy. 

The trace anomaly may, in general,  contain some total derivatives which originate  from the local counter terms that could be added to the effective action.
These terms thus are (regularization) scheme dependent and are not universal. The entanglement entropy, on the other hand, is a quantity which appears to be independent of the   
chosen regularization.  A natural question then arises: whether these total derivatives produce any non-trivial contribution to the entropy?

In this note we analyze this question. In dimension $d=4$ the only total derivative which may contribute to the trace anomaly is $\Box R$. It originates from $R^2$
in the effective action. In dimension $d=6$ there is much more freedom and there appears a set of possible terms. Generally they are rather complicated for the analysis.
Here, for the purposes of simplicity, we focus   on some particular terms, $\Box R^2$ and $\Box(R_{\mu\nu}R^{\mu\nu})$, which appear to be among the simplest ones. 
In all these cases the gravitational entropy is non-vanishing. In particular, for the term $\Box(R_{\mu\nu}R^{\mu\nu})$, the entropy is non-zero even if the surface has no extrinsic curvature. This is especially interesting in the light of the discrepancy first found in \cite{Hung:2011xb}. We suggest that this discrepancy may
originate from the total derivative terms in the trace anomaly that are normally ignored in the entropy calculation.
Below we present our analysis.

\section{ Regularization method}
First we want to explain our method. Consider a general class of metrics of the type 
\be
ds^2=e^{2\sigma(x,r)}\left(dr^2+r^2d\tau^2\right)+(h_{ij}(x)+2K^a_{ij}(x)n^ar+..)dx^idx^j\, ,
\lb{1}
\ee
where $n^1=\cos\tau$ and $n^2=\sin\tau$.
The entangling surface $\Sigma$ is defined by condition $r=0$, $h_{ij}(x)$ is the intrinsic metric on $\Sigma$ and $K^a_{ij}(x), a=1,2$ is the extrinsic curvature of the surface.
Now, in order to compute the  entropy associated to surface $\Sigma$ we use the replica trick, for a review see \cite{Solodukhin:2011gn}. It consists in making a 
periodicity $2\pi n$, $n$ is an integer,  for the coordinate $\tau$ and then taking the limit $n\to 1$. This procedure introduces an angle deficit $2\pi (1-n)$ and thus produces a conical singularity.  The entropy then is obtained by differentiating the gravitational action with respect to $(n-1)$ and taking the limit $n\to 1$.
If extrinsic curvature of $\Sigma$
is non-vanishing then the singularity is  the squashed conical singularity studied in \cite{Fursaev:2013fta}.
In order to compute the corresponding curvature invariants we introduce some regularization. This regularization consists of two parts. First, we smooth the conical singularity
by replacing $g_{rr}\to g_{rr}f_n(r)$ with the regularization function
\be
f_n(r)=\frac{r^2+b^2n^2}{r^2+b^2}\, ,
\lb{2}
\ee
where $b$ is the regularization parameter later to be taken to zero.  This regularization was introduced in  \cite{Fursaev:1995ef} so that we shall call it FS regularization.
It should be applied every time we have a conical singularity. If the conical singularity is squashed (i.e. the extrinsic curvature of $\Sigma$ is non-vanishing)
FS regularization alone   does not lead to everywhere regular space with a finite curvature. Thus it should be supplemented by yet another regularization:
replace $ K^a_{ij}(x)n^ar$  by  $K^a_{ij}(x)n^ar^n$ in the metric. 
We stress that the terms in the metric that remain there if $\Sigma$ is a Killing horizon should not be regularized. Otherwise we would get deviations from Wald's entropy calculation
even for the Killing horizons.   
This second regularization is introduced in \cite{Fursaev:2013fta} and we shall call it FPS regularization. It should be used if the singular surface has a non-trivial extrinsic curvature.  The final result for the entropy  thus can be considered as coming from both  FS and FPS regularizations.
 . 

\section{ Regularized metric and scalar curvature}
With these explanations we consider the following regularized metric
\be\label{metric}
ds^2=e^{2\sigma(x,r)}[f_n(r)dr^2+r^2d\tau^2]+g_{ij}(x,r,\tau)dx^idx^j ,
\lb{3}
\ee
where
\be
&&\sigma(x,r)=\sigma_0(x)+\half\sigma_2(x)r^2+\cdots\ , \nonumber \\
&&g_{ij}(x,r,\tau)=h_{ij}(x)+2K_{ij}^a(x)n^ar^n+(K^aK^b)_{ij}n^an^br^{2n}+g_{ij}^{(2)}(x)r^2+\cdots\ .
\lb{4}
\ee
According to our prescription in (\ref{4}) we changed the power of $r$ only for terms which are due to
extrinsic curvature keeping the power of $r$ in all other terms unchanged. 

If $n=1$ we have the following relations for the metric (\ref{3}):
\be
&&\sigma_2= -\frac{1}{4}e^{2\sigma_0}(R_{abab}+2(\nabla_\Sigma\sigma_0)^2)\, ,\nonumber \\
&&g^{(2)}_{ij}=-\frac{1}{2}e^{2\sigma_0}R_{ajaj}-e^{2\sigma_0}(\nabla_i\nabla_j\sigma_0+\nabla_i\sigma_0 \nabla_j\sigma_0)\, ,\nonumber \\
&&\Tr g^{(2)}=  \frac{1}{2}e^{2\sigma_0}(R_{abab}-R_{aa}-2\Delta_\Sigma\sigma_0-2(\nabla_\Sigma\sigma_0)^2)\, .
\lb{5}
\ee

If $n\neq 1$ the  scalar curvature of regularized metric (\ref{4}) reads
\be
&&R=A_1+A_2f_n^{-1}(r)-A_3r^{-1}\partial_r f_n^{-1}(r)+A_4C_1(r)r^{n-2}+A_5C_2(r)r^{2(n-1)}\nonumber \\
&&+A_6C_3(r)r^{(n-1)}+A_7r^{2(n-1)}\, , 
\lb{6}
\ee
where we introduced
\be
&&A_1=R_\Sigma-4\Delta_\Sigma\sigma_0-6(\nabla_\Sigma\sigma_0)^2\, , \ \ A_2=-4(\sigma_2+\Tr g^{(2)})e^{-2\sigma_0}\, , \nonumber \\
&&A_3=e^{-2\sigma_0}\, , \ \ A_4=e^{-2\sigma_0}\Tr K^a n^a\, , \ \ A_5=e^{-2\sigma_0}\Tr (K^aK^b) n^an^b\, , \nonumber \\
&&A_6=e^{-2\sigma_0}\Tr K^a\Tr K^b n^an^b\, , \ \ A_7=-e^{-2\sigma}[(\Tr K)^2-\Tr K^2]\, , \nonumber \\
&&C_1(r)=2(1-{n^2}/{f_n(r)})+{n\, r\partial_rf_n(r)}f^{-2}_n(r)\, ,\nonumber \\
&&C_3(r)=(1-n^2/f_n(r))\, , \ \ \  C_2(r)=-C_1(r)-C_3(r)\, .
\lb{7}
\ee
Imposing $n=1$ in (\ref{6}) we find  the Gauss-Codazzi relation
\be
R=R^\Sigma-4\sigma_2e^{-2\sigma_0}-4\Delta_\Sigma\sigma_0-6(\nabla_\Sigma\sigma_0)^2-4e^{-2\sigma_0}\Tr g^{(2)}-e^{-2\sigma_0}[(\Tr K)^2-\Tr K^2]\, .
\lb{7}
\ee
With the help of relations (\ref{5}) it takes the usual form
\be
R=R_\Sigma+2R_{aa}-R_{abab}+\Tr \hat{K}^2-(\Tr \hat{K})^2\, ,
\lb{8}
\ee
where $\hat{K}^a_{ij}=e^{-\sigma_0}K^a_{ij}, \ a=1,2$ is the extrinsic curvature.

For the square root of the determinant of metric (\ref{4}) we find that
\be
&&\sqrt{g}=\sqrt{h}\left(1+B_1 r^n+B_2r^{2n}+B_3r^2+..\right)\, , \nonumber \\
&&B_1=\Tr K^a n^a\, , \ \ B_2=\frac{1}{2}(\Tr K^a\Tr K^a-\Tr (K^aK^b))n^an^b\, , \ \ B_3=\frac{1}{2}\Tr g^{(2)}\, .
\lb{8-1}
\ee

\section{ Integrals over a squashed cone}
In this section we want to compute the contribution of a total derivative due to a conical singularity. In  fact, an obvious geometric quantity which is a total derivative is the
scalar curvature in two dimensions.  This was the first case analyzed in \cite{Fursaev:1995ef} in order to illustrate the distributional nature of the curvature due to a conical singularity. The procedure considered in \cite{Fursaev:1995ef} was the following. Let us first take a disk of a fixed radius $r_0$ in the plane $(r,\tau)$. Then we consider the integral
of the scalar curvature $R$ for the FS regularized metric. Formally  the radial  integral can be taken from $r=0$ to $r=r_0$. However, the term at $r=0$ vanishes for the FS regularized metric.
So that only term at $r=r_0$ is important. One decomposes this term in powers of $(1-n)$ to linear order and then takes the limit when the regularization parameter  $b\rightarrow 0$ provided the value of $r_0$ is kept fixed. The result of this procedure is finite and (for small $r_0$) independent of $r_0$.

Now we   want to repeat this procedure and compute the   integral of $\Box R$ over a regularized squashed cone and extract the contribution which is due to the
conical singularity. As above we consider a disk of radius $r_0$, where $r_0$ is small but finite. Then the integral reduces to two boundary terms, at $r=0$ and $r=r_0$,
\be
\int_{{\cal M}_n}\Box R=\int_{r=r_0} \frac{\sqrt{g}}{\sqrt{f_n}}r\partial_r R-\int_{r=0} \frac{\sqrt{g}}{\sqrt{f_n}}r\partial_r R\, ,
\lb{10}
\ee
where integration goes over $\tau$ (from $0$ to $2\pi n$) and $x^i$, and $g$ is determinant for the metric (\ref{4}), see eq. (\ref{8-1}). We notice that for curvature (\ref{6}), in the limit of small $r$ and provided that $b$ is
kept finite and $n>1$, we have that $r\partial_r R\sim r^{2n-2}$ vanishes at $r=0$.  Thus there is no ``internal boundary'' in (\ref{10}) and the integral reduces to the boundary term at $r=r_0$.
We expand the first term in  (\ref{10}) in powers of $(1-n)$ and then take the limit $b\to 0$ while keeping $r_0$ small but finite. 
The result of this procedure is
\be
\frac{\sqrt{g}}{\sqrt{f_n}}r\partial_r R=\frac{4(n-1)A_4}{r_0}+(4A_4B_1+2A_7)(n-1)+..\, ,
\lb{x1}
\ee
where we neglect the terms which are either higher powers of $(n-1)$ or of the regularization parameter $b$.  

Integrating over $\tau$ we use
\be
\int_0^{2\pi n}d\tau n^a=O(n-1)\, , \ \ \int_0^{2\pi n}d\tau n^a n^b =\pi \delta^{ab}+O(n-1)\, .
\lb{x2}
\ee
Therefore,  the first term in (\ref{x1}) integrated over $\tau$ is of the second order in $(n-1)$ while the second term gives (where we include the integration over $x$)
\be
\int_\Sigma\int_0^{2\pi n}\frac{\sqrt{g}}{\sqrt{f_n}}r\partial_r R=4\pi(n-1)\int_\Sigma (A_7+\Tr\hat{K}^a\Tr\hat{K}^a)=4\pi(n-1)\int_\Sigma \Tr\hat{K}^2\, .
\lb{x3}
\ee
The further integration over $x$ gives us the following result
\be
\int_{{\cal M}_n}\Box R=n\int_{{\cal M}_{n=1}}\Box R+ 4\pi (n-1)\int_\Sigma \Tr \hat{K}^2\, .
\lb{11}
\ee
Similarly, for the integral of $\Box R^2$ we find that
\be
\int_{{\cal M}_n}\Box R^2=n\int_{{\cal M}_{n=1}}\Box R^2 +8\pi (n-1)\int_\Sigma R\, \Tr \hat{K}^2\, ,
\lb{12}
\ee
where  in the r.h.s. of this equation the scalar curvature $R$ takes the form (\ref{7}) (or, equivalently, (\ref{8})).
Interestingly, the surface term in (\ref{11}) 
is non-zero even if the spacetime is flat. This makes it similar (but not identical) to the famous Gibbons-Hawking term.

The above analysis is based on the form of the metric (\ref{3})-(\ref{4}). This metric for $n>1$, if extended to infinite values of $r$, may have large
deviations from the original metric ($n=1$).  This deviation may be a reason for concern whether (\ref{3})-(\ref{4}) is a well-defined regularization.
Although we can not exclude the existence of other regularizations, the consequences of which should be further investigated,
here we would like to argue that (\ref{3})-(\ref{4}) is a legitimate form for the conical metric. First of all, for a disk of any finite radius $r_0$
the difference $(n-1)$ can be made arbitrary small so that the ``large'' deviation never occurs, the $(n-1)$ deformation of the metric would be made arbitrary small
in any appropriate norm. Moreover, the increasing radius $r_0$ does not change the $K^2$-term in (\ref{11}).

 The other reasoning in favor of  (\ref{3})-(\ref{4}) is the following. For $n=1$ the metric can be expressed in terms of complex variable
$z=re^{i\tau}$ (and $\bar{z}=re^{-i\tau}$). The part of the metric which linearly depends on the extrinsic curvature takes the form of a sum of two  holomorphic functions, of $z$ and $\bar{z}$ respectively.
When the periodicity of $\tau $ is $2\pi n$ one can redefine $\tau=n\phi$. Then it is natural to demand that for any integer $n$ the conical metric remains to be 
holomorphic of variable $z=re^{i\phi}$ (and $\bar{z}=re^{-i\phi}$). In this way the conical metric for integer $n>1$ shares same analytical properties as the metric
for $n=1$. This condition uniquely fixes the structure of the metric (\ref{3})-(\ref{4}). From this point of view  the replacement  $r\to r^n$ in (\ref{3})-(\ref{4}) should be viewed not as a regularization procedure but as a definition of the conical metric for an integer $n>1$.

\section{Holographic entanglement entropy}
As  first application of our finding we consider the generalization of the holographic proposal \cite{Ryu:2006ef} for the entanglement entropy.
In the holographic duality the AdS gravity may be described by an action which includes terms quartic in derivatives. The general structure of such an action then includes also a
total derivative term, 
\be\lb{hf2}
I=-\int_{{\cal M}^{(d+1)}}\sqrt{g}d^{d+1}x~
\left[\frac{R}{ 16\pi G_{(d+1)}}+2\Lambda+ \lambda_1 R_{\mu\nu\alpha\beta}R^{\mu\nu\alpha\beta}+\lambda_2R_{\mu\nu}R^{\mu\nu}+
\lambda_3R^2+\lambda_4\Box R\right] .
\ee
Respectively, the generalized holographic entropy is a combination of the proposal made in \cite{Fursaev:2013fta} and our finding (\ref{11})
\be\lb{hf5}
S({\cal H})=\frac{A(\cal H) }{ 4G_{(d+1)}}+
4\pi \int_{\cal H} \left[
2\lambda_1 (R_{ijij}-\tr k^2)    +\lambda_2 
(R_{ii}-\frac{1}{2} k^2)+   2\lambda_3R  -\lambda_4\tr k^2)      \right]\, ,
\ee
where $\cal H$ is  a co-dimension 2 surface which bounds the entangling surface $\Sigma$, $k$ is the extrinsic curvature of $\cal H$.
The surface $\cal H$ is supposed to be a minimizer of the functional (\ref{hf5}). 
If $d=4$ the surface $\cal H$ has dimension three and the holographic entropy (\ref{hf5}) is supposed to reproduce the entanglement entropy of 
a conformal field theory with general conformal charges.

\section{Conformal anomaly and entanglement entropy in four dimensions}
In four dimensions the trace anomaly is a combination of the following terms
\be
&&\la T\ra=-aE_4+b W^2+c\Box R\, ,\nonumber \\
&&E_4=R_{\alpha\beta\mu\nu}R^{\alpha\beta\mu\nu}-4R_{\mu\nu}R^{\mu\nu}+R^2\, ,\nonumber \\
&&W^2=R_{\alpha\beta\mu\nu}R^{\alpha\beta\mu\nu}-2R_{\mu\nu}R^{\mu\nu}+\frac{1}{3}R^2\, ,
\lb{13}
\ee
where  $W$ is the Weyl tensor  and $E_4$ is the Euler density in four dimensions.  The first two terms in (\ref{13}) are universal. They come from a conformal variation of the non-local part of the CFT effective action. On the other hand, the last term originates from a conformal variation of a local term $R^2$ which could be added
 to the effective action. This term depends on the regularization scheme and it is not universal. Respectively, the $c$-term in (\ref{13}) is not universal.
 
On the other hand, the trace anomaly integrated over a conical space ${\cal M}_n$
\be
\int_{{\cal M}_n}\la T\ra= n\int_{{\cal M}_{n=1}}\la T\ra+(1-n)s_0/2 \, ,
\lb{14}
\ee
is related to the logarithmic term in the entanglement entropy,
\be
S=\frac{N_sA(\Sigma)}{48\pi\epsilon^2}+s_0\ln \epsilon\, .
\lb{15}
\ee
The term $s_0$ in (\ref{14}) is  then given by a surface integral
\be
s_0=16\pi\int_\Sigma \left(aR_\Sigma-bK_\Sigma +\frac{c}{2} \tr \hat{K}^2\right)\, ,
\lb{16}
\ee
where $R_\Sigma$ is intrinsic curvature of surface $\Sigma$,  and we define
\be
K_\Sigma=R_{abab}-R_{aa}+\frac{1}{3}R-(\Tr \hat{K}^2-\frac{1}{2}(\Tr \hat{K})^2)\, .
\lb{17}
\ee
The $a$- and $b$-contributions to logarithmic term (\ref{16}) have been obtained earlier, see  \cite{Ryu:2006ef} and \cite{Solodukhin:2008dh}.

\section{Some puzzles}
The $c$-term in (\ref{16}) is new. The existence of this term is a direct consequence of (\ref{11}). However, its presence in the logarithmic term of entanglement entropy
is rather puzzling. In a conformal field theory
$s_0$  is expected to be conformally invariant  and indeed the $a$- and $b$-terms in (\ref{16}) are conformal invariants. However, the $c$-term is not invariant under
conformal transformations. Indeed, under Weyl rescaling $g_{\mu\nu}\rightarrow g_{\mu\nu}e^{-2\omega}$, $n^a_\mu\rightarrow e^{-\omega}n^a_\mu$ it changes as
\be
\int_\Sigma \Tr \hat{K}^2\rightarrow \int_\Sigma \Tr \hat{K}^2+\int_\Sigma \Tr \hat{K}^an_a^\alpha\partial_\alpha \omega\, 
\lb{17-1}
\ee
and is invariant only if normal derivative of $\omega$ vanishes on $\Sigma$. So that the $c$-term breaks conformal invariance down to transformations which preserve the
extrinsic curvature.

 Moreover, for a sphere $\Sigma=S_2$ in flat spacetime, we have that 
\be
R_\Sigma=\Tr \hat{K}^2 \, ,
\lb{18}
\ee
so that the $c$-term takes exactly same form as the $a$-term. In  this case the $b$-term disappears and the whole contribution to the
logarithmic terms is due to the  Euler number of the sphere multiplied by $(2a+c)$,
\be
s_0(S_2)=32\pi^2(2a+c)\, .
\lb{18-1}
\ee
 Usually one considers the logarithmic term in the entropy for a round sphere as
a simple way to identify the $a$-charge of the CFT (see for instance \cite{Casini:2011kv}, \cite{Bhattacharyya:2013gra}). However, we see that, in general, this term may also have  a part that depends on $c$.

The dependence of entanglement entropy on $c$ should mean that the entropy depends on the regularization. So far no indication of such a dependence has been found\footnote{However, we were informed by Christopher Eling about his earlier unpublished work \cite{Cris} on resolution of the mismatch in the logarithmic term in the entropy of gauge fields first observed by Dowker \cite{Dowker:2010bu}. Eling uses the previous results of  \cite{Brown:1986jy}. According to  \cite{Brown:1986jy} in the trace anomaly due to the  gauge spin-1 field the parameter $c$ is non-zero. Using these results Eling obtains $(4a+2c)\ln\epsilon$ for the logarithmic term in the entanglement entropy in agreement with our eq.(\ref{18-1}) (he uses different normalization for $a$ and $c$). Then, for $a=62n_1/360$ and $c=-n_1/6$ as in \cite{Brown:1986jy} he finds that $a+c/2=32n_1/360$ in agreement with Dowker.
 This relates the mismatch to the presence of $c$-term in the trace anomaly.
}
 neither     in direct lattice type calculations of the entropy nor in the numerous holographic calculations that use the prescription of Ryu-Takayanagi \cite{Ryu:2006ef}.
The absence of $c$-term in the holographic analysis however may have a simple explanation. As shows the analysis in original paper  \cite{Henningson:1998gx} the total derivative term  vanishes ($c=0$) in the holographic trace anomaly in four dimensions. However, the total derivatives appear in the holographic trace anomaly in six dimensions.

\section{Earlier observed discrepancy in six dimensions}
Curiously enough, paper \cite{Hung:2011xb} does observe some discrepancy between the holographic calculation of entanglement entropy (using Jacobson-Myers functional for the holographic minimal surface) and the CFT trace anomaly calculation
if one uses Wald's prescription to compute the entropy in $d=6$.  This discrepancy has not yet been  explained in the literature. As observed in \cite{Hung:2011xb} the discrepancy comes from the conformal invariant $I_3$, the only conformal invariant in $d=6$ which contains total derivatives, and
it is conceivable that it originates entirely from the total derivative terms present in $I_3$. $\Box R^2$ (\ref{12}) is one of such terms. The discrepancy found in  \cite{Hung:2011xb} is for surfaces without $O(2)$ symmetry but
with vanishing extrinsic curvature. So that (\ref{12}) can not give the required contribution to explain the discrepancy. However, there are more total derivative terms
in $I_3$ (see for instance  \cite{Bastianelli:2000hi})  and some of them may have the required properties. 
In support to these expectations we shall consider a particular example of the total derivative which does appear in the trace anomaly in six dimensions and has  the required property.

\section{More general metric  }
First we need to generalize the metric (\ref{3}), (\ref{4}). Indeed, it was assumed in (\ref{3}), (\ref{4}) that to second order in $r$ the $\tau$-dependence 
of the metric may appear only due to the extrinsic curvature $(K^aK^b)n^an^b\, r^2$ so that $g^{(2)}$ does not depend on $\tau$. This,  however, is not the most general
situation. Indeed, in the examples considered in \cite{Hung:2011xb} the extrinsic curvature of the entangling surface is zero. However the surface is not $O(2)$ invariant
in the transverse subspace due to $\tau$-dependent terms in the $r^2$ order of the metric when expanded near the surface.

Motivated by these examples   we consider the following generalization of the (not yet regularized)  metric
\be
&&ds^2=e^{2\sigma(x,r)}[dr^2+r^2d\tau^2]+g_{ij}(x,r,\tau)dx^idx^j ,
\lb{19} \\
&&g_{ij}(x,r,\tau)=h_{ij}(x)+2K_{ij}^a(x)n^ar+r^2((K^aK^b)_{ij}n^an^b+H^{ab}_{ij}(x)n^an^b )+\cdots\ , \nonumber \\
&&\sigma(x,r)=\sigma_0(x)+\half\sigma_2(x)r^2+\cdots \, \, .\nonumber 
\ee
This metric is obviously $2\pi n$ periodic if $n$ is an integer.  It should be noted that the trace part of $H^{ab}_{ij}$ in (\ref{19}) is identical to what we called
$g^{(2)}_{ij}(x)$ in metric (\ref{4}).
The transverse components of the Ricci tensor of this metric read
\be
&&R_{rr}=-2\sigma_2-e^{2\sigma_0}[\Delta_\Sigma\sigma_0+2(\nabla_\Sigma\sigma_0)^2]-\Tr H^{ab}n^an^b\, ,\lb{21} \\
&&r^{-2}R_{\phi \phi}=-2\sigma_2-e^{2\sigma_0}[\Delta_\Sigma\sigma_0+2(\nabla_\Sigma\sigma_0)^2]+\Tr H^{ab}n^an^b-\Tr H^{aa} \, , \nonumber\\
&&R_{r\phi}= \Tr H^{ab}n^a\epsilon_{bc}n^cr\, .\nonumber
\ee
In what follows we assume that the extrinsic curvature of the surface vanishes, $K^a_{ij}=0$, but the term $H^{ab}_{ij}$ is non-vanishing.
So that vector $\xi=\partial_\tau$ is locally a Killing vector, ${\cal L}_{\xi}g_{\mu\nu}=O(r^2)$. Thus, the surface at $r=0$ in the metric (\ref{19})
is some sort of generalized horizon. 
In this case the conical singularity does not appear to be ``squashed'' although it is not $O(2)$ invariant either.

If we use the FS regularization only, i.e. replace $g_{rr}\rightarrow f_n(r)g_{rr}$, then the regularized metric has everywhere finite 
curvature. However, the gradient of the curvature is divergent\footnote{We thank Joan Camps for pointing this out to us.} at $r=0$. 
Therefore, we need to use additionally the FPS regularization in order to make the derivatives of the curvature finite. The analysis shows that the divergence  is due to the traceless 
part of $H^{ab}$ in the metric (\ref{19}). Therefore, only this part needs to be regularized while the part of the metric due to the trace of $H^{ab}$ is 
independent of $\tau$ and thus it does not need to be regularized. This is the prescription advocated in \cite{Camps:2013zua}. With this prescription we replace
\be
H^{ab}_{ij}(x)n^an^b r^2\rightarrow \frac{1}{2}H_{ij}(x) r^2+(H^{ab}_{ij}(x)-\frac{1}{2}\delta^{ab}H_{ij}(x))n^an^b r^{2n}\, ,
\lb{19-2}
\ee
where $H_{ij}(x)=H^{ab}_{ij}\delta^{ab}$, in the metric (\ref{19}). If the  traceless part of $H^{ab}$ vanishes then metric (\ref{19}) (provided $K^a_{ij}$ vanishes as well)
possesses the Killing symmetry  and describes a Killing horizon at $r=0$. For this metric Wald's calculation of entropy is applicable and we do not expect
any modifications of this calculation. This explains why we  did not modify the power of $r$ in front of $H_{ij}(x)$ in (\ref{19-2}). 

\section{Entropy calculation}
As an example of a total derivative term we shall  consider $\Box (R_{\mu\nu}R^{\mu\nu})$. The respective integral over the conical space then reduces to a boundary term at $r=r_0$,
\be
\int_{{\cal M}_n}\Box (R_{\mu\nu}R^{\mu\nu})=\int_{r=r_0} \frac{\sqrt{g}}{\sqrt{f_n}}r \partial_r(R_{\mu\nu}R^{\mu\nu})\, ,
\lb{22}
\ee
in which we have to expand in powers of $(n-1)$ and take the limit $b\to 0$.
The result of this calculation for the metric (\ref{19}) regularized as we just explained  is rather simple and it depends only on $H^{ab}_{ij}$,
\be
\int_{{\cal M}_n}\Box (R_{\mu\nu}R^{\mu\nu})=8\pi(n-1)\int_\Sigma \left(\Tr H^{ab}\Tr H^{ab}-\frac{1}{2}(\Tr H^{aa})^2\right)\, ,
\lb{23}
\ee
where the trace is defined with respect to intrinsic metric $h_{ij}(x)$ of the surface. Not surprisingly, the result (\ref{23}) depends only on the traceless part of
$H^{ab}$.

This can be re-written in terms of the Ricci tensor projected on the transverse 
subspace, $R_{ab}=R_{\mu\nu}n^\mu_a n^\nu_b$, where $n_a^\mu, \ a=1,2$ is a pair of normal vectors to $\Sigma$. For metric (\ref{19}) we have that
$n_1^r=e^{-\sigma_0(x)}$ and $n_2^\tau=r^{-1}e^{-\sigma_0(x)}$. Then, using (\ref{21}), we have that
\be
\int_{{\cal M}_n}\Box (R_{\mu\nu}R^{\mu\nu})=8\pi(n-1)\int_\Sigma (R_{ab}-\frac{1}{2}\delta_{ab}R_{cc})^2\, .
\lb{24}
\ee
Together with equations (\ref{11}) and (\ref{12}) this is our main result.

The  entropy which follows from (\ref{24}) is
\be
S=8\pi \int_\Sigma \left(R_{ab}^2-\frac{1}{2}(R_{aa})^2\right)\, .
\lb{24-1}
\ee
It has the structure that resembles the one proposed in \cite{Hung:2011xb}
in terms of the Weyl tensor. However, in order to see whether there is a complete agreement we need to know the respective entropy which comes from all possible total derivative terms that appear in the trace anomaly. Work in this direction is in progress.

\section{General expression for entropy and some tests}
We can advance a bit more in our attempt to explain the discrepancy of \cite{Hung:2011xb}.
It is clear from the analysis above that the possible contributions of the total derivative terms in the trace anomaly in six dimensions should be a combination of invariants constructed from matrix $H^{ab}_{ij}$. There are in general four such invariants so that the respective entropy is a linear combination 
\be
 S=\int_\Sigma\left(\alpha_1 \Tr H_{ab}\Tr H^{ab}+\alpha_2 (\Tr H_{aa})^2+\beta_1 H^{ab}_{ij}H^{ab,ij}+\beta_2 H^{aa}_{ij}H^{aa,ij}\right)\, .
\lb{25}
\ee
It is natural to expect that only the traceless part of $H^{ab}$ contributes to the missing entropy. Then we have $\alpha_2=-\alpha_1/2$ and $\beta_2=-\beta_1/2$.

In the examples considered in \cite{Hung:2011xb} the matrix $H^{ab}$ has only one non-vanishing component, $H^{11}$. Therefore among these four invariants there are only two independent
\be
 S=\int_\Sigma \left(\alpha (\Tr H^{11})^2+\beta H^{11}_{ij}H^{11,ij}\right)\, ,
\lb{26}
\ee
where $\alpha=\alpha_1+\alpha_2$ and $\beta=\beta_1+\beta_2$.

The analysis of \cite{Hung:2011xb} can be summarized as follows. The mismatch in the entropy that they have found takes the form
\be
\Delta S=-\pi B_3 g A(\Sigma)\, \ln \epsilon\, ,
\lb{27}
\ee
where $B_3$ is the central charge which corresponds to conformal invariant $I_3$.
In \cite{Hung:2011xb} they considered four cases (we use their notations and set all radii to 1):\\

\noindent {\bf a).} $R^1\times S^2\times S^3$ with $\Sigma= S^1\times S^3$. \cite{Hung:2011xb} finds  that in this case $g=6$. 
We find that $(\Tr H^{11})^2=1$ and $H^{11}_{ij}H^{11,ij}=1$ for this geometry. \\

\noindent {\bf a').} $R^1\times S^2\times S^3$ with $\Sigma=S^2\times S^2$. \cite{Hung:2011xb} finds $g=8$. Respectively we find that $(\Tr H^{11})^2=4$ and $H^{11}_{ij}H^{11,ij}=2$.\\\

\noindent {\bf b).} $R^3\times S^3$ with $\Sigma=S^2\times R^2$. \cite{Hung:2011xb} finds $g=8$. We find $(\Tr H^{11})^2=4$ and $H^{11}_{ij}H^{11,ij}=2$. \\

\noindent {\bf c).} $R^2\times S^4$ with $\Sigma=S^3\times R^1$. \cite{Hung:2011xb} finds $g=6$. We find $(\Tr H^{11})^2=9$ and $H^{11}_{ij}H^{11,ij}=3$.

\bigskip

According to our proposal,  $g A(\Sigma)$ should be identified with (\ref{25})-(\ref{26}). There are two observations which may serve  as some tests on this proposal.
First, the cases {\bf a')} and {\bf b)} are characterized by same $H$-invariants. Therefore, if we are right then we expect that their entropy mismatch should be the same.
This is indeed the case! Then, the three independent  cases give us three equations on parameters $\alpha$ and $\beta$:
\be
&&\alpha+\beta=6\, ,\nonumber \\
&&4\alpha+2\beta=8\, ,\nonumber \\
&&9\alpha+3\beta=6\, .
\lb{28}
\ee
The first two equations have solution: $\alpha=-2$ and $\beta=8$. With these values, the third equation in (\ref{28}) holds automatically! This is second
test on our proposal.

In fact, provided the missing entropy (\ref{26}) depends only on the traceless part $\tilde{H}^{ab}_{ij}=H^{ab}_{ij}-\frac{1}{2}\delta^{ab} H^{cc}_{ij}$ we  are now able to get the complete expression
\be
S=\int_\Sigma\left(-4\Tr \tilde{H}^{ab}\Tr \tilde{H}^{ab}+16 \tilde{H}^{ab}_{ij}\tilde{H}^{ab,ij}\right)\, .
\lb{28-1}
\ee

It should be noted however that the full resolution of the discrepancy may be a rather complicated problem.
The reason is the following. The total derivatives may appear on both sides of the holographic relation.
On the CFT side they appear in the holographic trace anomaly as derived in \cite{Henningson:1998gx}, \cite{Bastianelli:2000hi}.
On the other hand, this should be compared to the holographic entropy which itself may be modified by the presence of the total derrivative
terms in the AdS gravitational action. An example of this modification we have seen  in section 5. If terms of  6th order in derivative are allowed,
the structure of possible total derivative terms is much richer and some of them may produce  contributions to the holographic entropy
that do not disappear even if the surface is minimal.  Since not all these contributions are at the moment known we can not yet bring together all pieces of the puzzle and
fully resolve the problem. 

On the other hand, our proposal (\ref{25}), (\ref{28-1})  is perhaps an easier way to attack the problem. It represents the total missing entropy which may come from both sides of the holographic relation.
It is simple and can be easily checked for new examples of surfaces for which the  mismatch in the entropy is found.

\section{Conclusions}
We have analyzed  the possibility that the total derivative terms in the gravitational action may lead to some non-trivial contributions to the entropy.
Rather surprisingly, in the examples of total derivative terms which we consider in this note we have found that such a contribution does exist. 
 This observation may have many applications. We have briefly discussed the relevance of our finding to the logarithmic term in entanglement entropy 
and its relation to the conformal anomaly. In four dimensions the anomaly may in general contain a total derivative which originates from a local term in the quantum effective action.
Then we predict that this term would contribute to the logarithmic term in a particular way which sometimes (for any sphere in Minkowski spacetime) mimics the contribution of the $a$-charge.

In six dimensions the structure of total derivative terms in the trace anomaly is much richer. We analyze some of them. In particular, we have found that
they may produce  contributions which do not disappear when the surface has no extrinsic curvature.  The corresponding entropy is not of Wald's type.
We suggest that the entropy which comes from the total derivative terms in the trace anomaly is the source for  the discrepancy between the holographic entanglement entropy and
Wald's entropy earlier observed in \cite{Hung:2011xb}. In a wider context this should mean that the total derivative  terms in the gravitational action can not be
neglected and may lead to some non-trivial gravitational entropy. This entropy may manifest itself for time-dependent metrics with a generalized horizon.
Further implications for the thermodynamics of horizons of this type are worth exploring.

These conclusions are made under assumption of the correspondence between the gravitational action  and the entropy as suggested by the application of the conical singularity method.
It would be interesting to verify in an independent way whether the predicted contributions 
(such as $c$-term in four dimensions)  do appear in entanglement  entropy. The negative answer to this question would possibly impose certain restrictions on the entropy/action correspondence.
These restrictions (once identified) would provide us with the useful information on the applicability of the method of conical singularity.

\section*{Acknowledgements} 
A.F.A. and A.P. would like to thank  Laboratoire de Math\'ematiques et Physique Th\'eorique (LMPT) for kind hospitality where part of this work was done.
A.P. acknowledges the support from the Metchnikov's post-doctoral program of the Embassy of France  in Russia and the support from the RFBR grant 13-02-00950.
A.P. is also grateful to the Mainz Institute for Theoretical Physics (MITP) for its hospitality and its partial support on the initial stage of this work.
S.S. thanks Misha Smolkin for  inspiring discussions on earlier stages of this project. We thank Joan Camps and Cristopher Eling for useful comments.

\newpage
\appendix
\section{Appendix}
Here we would like to discuss an alternative way to compute (\ref{10}) as a bulk integral. It was claimed \cite{Dong} that  the bulk method 
gives zero result for the entropy due to $\Box R$. So that our purpose is to examine this possibility.
For simplicity we consider a simple metric
\be
ds^2=f_n(r)dr^2+r^2d\tau^2+(1+K_1 r^n\cos(\tau)+K_2r^n\sin(\tau))^2(dy^2+dz^2)\, .
\lb{a1}
\ee
For this metric we find  after the integration over $\tau$ and replacing $r=bx$ that
\be
\int_{{\cal M}_n}\Box R=\int_\Sigma \int_0^{r_0/b} dx \, 4\pi(1-n)\, b^{2n-2} x^{2n-3} \,   Q_n(x)
 \frac{(K_1^2+K_2^2)}{(x^2+1)^{1/2}(x^2+n^2)^{9/2}}
 \lb{a2}
 \ee
 where we skipped the terms proportional to $\sin(n\pi)$ which do not contribute to the entropy and we defined
 \be
 &&Q_n(x)= 2\,{x
}^{10}+11\,{x}^{8}+10\,{x}^{6}+2\,n\,{x}^{10}-4\,{n}^{2}{x}^{4}+6\,{x}
^{6}{n}^{7}+2\,{x}^{4}{n}^{9}+2\,{x}^{10}{n}^{3}+6\,{x}^{8}{n}^{5}\nonumber \\
&&-63
\,{x}^{6}{n}^{4}-13\,{x}^{4}{n}^{6}+16\,{n}^{7}{x}^{2}-18\,{n}^{6}{x}^
{2}-29\,{x}^{6}{n}^{3}-17\,{x}^{6}{n}^{2}-56\,{x}^{4}{n}^{4}+26\,{x}^{
6}n\nonumber \\
&&-16\,{x}^{4}{n}^{3}+24\,{x}^{8}n-9\,{x}^{8}{n}^{4}-8\,{x}^{8}{n}^{3
}-20\,{x}^{8}{n}^{2}-6\,{x}^{10}{n}^{2}+6\,{n}^{9}{x}^{2}\nonumber \\
&&+3\,{x}^{4}{n
}^{8}+23\,{x}^{4}{n}^{7}+7\,{x}^{6}{n}^{5}+{x}^{4}{n}^{5}+4\,{n}^{9}-4
\,{n}^{8} \, .
\lb{a3}
\ee
Notice that  since the integration over $r$ goes from $0$ to $r_0$ the respective integration for $x$ goes from $0$ to $r_0/b$.
In the limit $b\rightarrow 0$ the latter goes to infinity and it seems that one could replace the upper limit in (\ref{a2}) by infinity.
Below we examine this replacement which appears to be a tricky one. 

First,  we consider the decomposition of function $Q_n(x)$ to linear order in $(1-n)$,
\be
Q_n(x)=4\,{x}^{8}-60\,{x}^{6}-60\,{x}^{4}+4\,{x}^{2}+ (4-4\,{x}^{10} +a_1 x^2 +a_2 x^4+a_3 x^6 +a_4 x^8)(n-1)\, .
\lb{a4}
\ee
The exact values of $a_k$, $k=1,2,3,4$ are not important since the respective terms will not contribute to the entropy.

Suppose that we extend the integration over $x$ in (\ref{a2}) till infinity. Then for the individual terms the integration gives us
\be
I_k=\int_0^\infty \frac{dx\, x^{2n-3+k}}{(x^2+1)^{1/2}(x^2+n^2)^{9/2}}=\frac{\pi}{48}\frac{(k-2)(k-4)(k-6)(k-8)}{16\sin(\frac{\pi k}{2})}+O(n-1)\, .
\lb{a5}
\ee
This formula is valid if $k\neq 0,\, 10$.
In particular, we find that
\be
I_2=I_8=\frac{1}{8}\, , \, \, I_4=I_6=\frac{1}{24}\, .
\lb{a6}
\ee
This allows us to compute the contribution to the integral (\ref{a2}) of the $(n-1)^0$ terms in (\ref{a4}).  This contribution is $16\pi(n-1)$.

Now let us consider the terms of order $(n-1)$ in (\ref{a4}). The two terms, $4-4x^{10}$, may contribute to the entropy. Indeed, 
the integral $I_k$ (\ref{a5}) may contain a pole $\sim 1/(n-1)$ if $k=0$ and $k=10$.  In these two cases, instead of (\ref{a5}) one uses \cite{Dong}
\be
&&I_0=\int_0^\infty \frac{dx\, x^{2n-3}}{(x^2+1)^{1/2}(x^2+n^2)^{9/2}}=\frac{1}{2(n-1)}+..\nonumber \\
&&I_{10}=\int_0^\infty \frac{dx\, x^{2n+7}}{(x^2+1)^{1/2}(x^2+n^2)^{9/2}}=-\frac{1}{2(n-1)}+..\, .
\lb{a7}
\ee
With these values for the integrals the terms of order $(n-1) $ in (\ref{a4}) contribute to the entropy  $-16\pi (n-1)$.  Together with the contribution
of $(n-1)^0$ terms in (\ref{a8}) this gives us $16\pi(n-1)-16\pi(n-1)=0$. So that its seems that using the bulk method one could conclude, as in \cite{Dong}, that
the integral (\ref{a2}) vanishes and hence there is no entropy for $\Box R$. This conclusion contradicts the boundary terms method discussed in the main text.
The source of this contradiction is in the replacement  of the upper limit $r_0/b$ in (\ref{a2}) with infinity.
In fact, this replacement is legitimate in the integrals (\ref{a5}), (\ref{a6}).  However we have to be more careful when evaluate the integrals (\ref{a7}).
In the integral $I_0$ the pole $1/(n-1)$ arises due to integration over small values of $x$, $\int dx x^{2n-3}=\frac{x^{2n-2}}{2(n-1)}$. The limit $x\to 0$ is well defined if $n>1$.
However, approaching $n=1$ from above the divergence of the integral manifests in the pole $1/2(n-1)$. 
On the other hand, the integral $I_{10}$ is divergent on the upper limit. This divergence is cured if $n<1$. Approaching $n=1$ from below this integral shows the pole
$-1/2(n-1)$ as in (\ref{a7}). Clearly, the two integrals, $I_0$ and $I_{10}$, can not be well defined for the same values of $n$. Assumption  that $n>1$  makes legitimate the 
extension of integration in $I_0$ till infinity. However, for $n>1$  the second integral, $I_{10}$, is divergent on the upper limit and hence this limit can not be extended to infinity.
Instead, we should integrate till $r_0/b$ as is originally defined in (\ref{a2}). Then we find
\be
I_{10}=\int_0^{r_0/b} \frac{dx\, x^{2n+7}}{(x^2+1)^{1/2}(x^2+n^2)^{9/2}}=\ln (r_0/b)+O(n-1)\, ,
\lb{a8}
\ee
where we assume  that $b$ is small but finite. Thus, in the careful treatment there is no a pole in the integral $I_{10}$. Summing up all contributions, the one from $(n-1)^0$ terms in (\ref{a4}) and the other from $(n-1)$ term in (\ref{a4}), we find $16\pi (n-1)-8\pi(n-1)=8\pi(n-1)$. 
Finally, we arrive at
\be
\int_{{\cal M}_n}\Box R=8\pi (n-1)\int_\Sigma (K_1^2+K_2^2)
\lb{a9}
\ee
in a complete agreement with (\ref{11}) since for metric (\ref{a1}) $\Tr K^2=2(K_1^2+K_2^2)$.


\newpage
\begin{thebibliography}{999}

{\frenchspacing \parskip=2mm

\bibitem{Wald:1993nt} 
  R.~M.~Wald,
  ``Black hole entropy is the Noether charge,''
  Phys.\ Rev.\ D {\bf 48}, 3427 (1993)
  
\bibitem{Jacobson:1993xs} 
  T.~Jacobson and R.~C.~Myers,
  ``Black hole entropy and higher curvature interactions,''
  Phys.\ Rev.\ Lett.\  {\bf 70}, 3684 (1993)
  [hep-th/9305016].
  
\bibitem{Jacobson:1993vj} 
  T.~Jacobson, G.~Kang and R.~C.~Myers,
  ``On black hole entropy,''
  Phys.\ Rev.\ D {\bf 49}, 6587 (1994)
  [gr-qc/9312023].

\bibitem{Fursaev:1995ef} 
  D.~V.~Fursaev and S.~N.~Solodukhin,
  ``On the description of the Riemannian geometry in the presence of conical defects,''
  Phys.\ Rev.\ D {\bf 52}, 2133 (1995)
  [hep-th/9501127]. \\
  D.~V.~Fursaev and S.~N.~Solodukhin,
  ``On one loop renormalization of black hole entropy,''
  Phys.\ Lett.\ B {\bf 365}, 51 (1996)
  [hep-th/9412020].
  
\bibitem{Lewkowycz:2013nqa} 
  A.~Lewkowycz and J.~Maldacena,
  ``Generalized gravitational entropy,''
  JHEP {\bf 1308}, 090 (2013)
  [arXiv:1304.4926 [hep-th]].
  
\bibitem{Barvinsky:1985an} 
  A.~O.~Barvinsky and G.~A.~Vilkovisky,
  ``The Generalized Schwinger-Dewitt Technique in Gauge Theories and Quantum Gravity,''
  Phys.\ Rept.\  {\bf 119}, 1 (1985).

\bibitem{Ryu:2006ef} 
S. Ryu and T. Takayanagi, 
``Holographic Derivation of Entanglement Entropy from AdS/CFT,'' 
Phys. Rev. Lett. {\bf 96} (2006) 181602
[hep-th/0603001].\\
  S.~Ryu and T.~Takayanagi,
  ``Aspects of Holographic Entanglement Entropy,''
  JHEP {\bf 0608}, 045 (2006)

\bibitem{Solodukhin:2008dh} 
  S.~N.~Solodukhin,
  ``Entanglement entropy, conformal invariance and extrinsic geometry,''
  Phys.\ Lett.\ B {\bf 665}, 305 (2008)
  [arXiv:0802.3117 [hep-th]].
  
    
\bibitem{Hung:2011xb} 
  L.~Y.~Hung, R.~C.~Myers and M.~Smolkin,
  ``On Holographic Entanglement Entropy and Higher Curvature Gravity,''
  JHEP {\bf 1104}, 025 (2011)
  [arXiv:1101.5813 [hep-th]].

\bibitem{Solodukhin:2011gn} 
  S.~N.~Solodukhin,
  ``Entanglement entropy of black holes,''
  Living Rev.\ Rel.\  {\bf 14}, 8 (2011)
  
\bibitem{Fursaev:2013fta} 
  D.~V.~Fursaev, A.~Patrushev and S.~N.~Solodukhin,
  ``Distributional Geometry of Squashed Cones,''
  Phys.\ Rev.\ D {\bf 88}, no. 4, 044054 (2013)
  [arXiv:1306.4000 [hep-th]].
  

\bibitem{Casini:2011kv} 
  H.~Casini, M.~Huerta and R.~C.~Myers,
  ``Towards a derivation of holographic entanglement entropy,''
  JHEP {\bf 1105}, 036 (2011)
  [arXiv:1102.0440 [hep-th]].


\bibitem{Bhattacharyya:2013gra} 
  A.~Bhattacharyya, M.~Sharma and A.~Sinha,
  ``On generalized gravitational entropy, squashed cones and holography,''
  JHEP {\bf 1401}, 021 (2014)
  [arXiv:1308.5748 [hep-th]].
  
\bibitem{Henningson:1998gx} 
  M.~Henningson and K.~Skenderis,
  ``The Holographic Weyl anomaly,''
  JHEP {\bf 9807}, 023 (1998)
  [hep-th/9806087].
  
\bibitem{Camps:2013zua} 
  J.~Camps,
  ``Generalized entropy and higher derivative Gravity,''
  JHEP {\bf 1403}, 070 (2014)
  [arXiv:1310.6659 [hep-th]].
  
\bibitem{Dong:2013qoa} 
  X.~Dong,
  ``Holographic Entanglement Entropy for General Higher Derivative Gravity,''
  JHEP {\bf 1401}, 044 (2014)
  [arXiv:1310.5713 [hep-th], arXiv:1310.5713].
  
  \bibitem{Cris} Cristopher Eling, unpublished.
  
\bibitem{Brown:1986jy} 
  M.~R.~Brown, A.~C.~Ottewill and D.~N.~Page,
  ``Conformally Invariant Quantum Field Theory in Static Einstein Space-times,''
  Phys.\ Rev.\ D {\bf 33}, 2840 (1986).
  
  \bibitem{Dowker:2010bu} 
  J.~S.~Dowker,
  ``Entanglement entropy for even spheres,''
  arXiv:1009.3854 [hep-th]\\
  R. Myers, M.  Smolkin and A.~Patrushev, unpublished; reported in A. Patrushev, `Topics in Field Theory',  PhD Thesis, 2012, http://ir.lib.uwo.ca/etd/864/ \\
  C.~Eling, Y.~Oz and S.~Theisen,
  ``Entanglement and Thermal Entropy of Gauge Fields,''
  JHEP {\bf 1311}, 019 (2013)
  [arXiv:1308.4964 [hep-th]].
  
\bibitem{Bastianelli:2000hi}
  F.~Bastianelli, S.~Frolov and A.~A.~Tseytlin,
  ``Conformal anomaly of (2,0) tensor multiplet in six-dimensions and AdS / CFT correspondence,''
  JHEP {\bf 0002} (2000) 013
  [hep-th/0001041].
  \bibitem{Dong}
  X.~Dong and R.-X. Miao,  private communication.
}


\end{thebibliography}
\end{document}